%% file: arxiv.tex
\documentclass[twocolumn,aps,superscriptaddress,longbibliography]{revtex4}
\usepackage{graphicx}

\usepackage[fleqn]{amsmath}
\usepackage{amssymb}
\usepackage[dvipsnames]{xcolor}
\usepackage[normalem]{ulem}
\definecolor{goodgreen}{rgb}{0.1,0.5,0}
\definecolor{goodred}{rgb}{0.7,0,0}
\usepackage{float}
\usepackage{multirow}

\usepackage{tikz}
\usetikzlibrary{arrows.meta, positioning}
\usetikzlibrary{backgrounds}

\usepackage{listings} 

\usepackage{array,mathtools,amssymb,booktabs}
\newcolumntype{C}{>{$}c<{$}}

\AtBeginDocument{%
  \heavyrulewidth=.08em
  \lightrulewidth=.05em
  \cmidrulewidth=.03em
  \belowrulesep=.65ex
  \belowbottomsep=0pt
  \aboverulesep=.4ex
  \abovetopsep=0pt
  \cmidrulesep=\doublerulesep
  \cmidrulekern=.5em
  \defaultaddspace=.5em
}

\usepackage[colorlinks,urlcolor=goodgreen,citecolor=blue,linkcolor=goodred]{hyperref}
\usepackage{bm}
\allowdisplaybreaks
\usepackage{todonotes} 
\usepackage{cleveref} 
\usepackage[separate-uncertainty=true]{siunitx}
\usepackage{physics} 
\usepackage{csquotes}

\usepackage{pgfplots} 

\usepackage{import}
\usepackage{xifthen}
\usepackage{pdfpages}
\usepackage{transparent}

\renewcommand{\vec}[1]{\bm{#1}}
\renewcommand{\epsilon}{\varepsilon}

\DeclareSIUnit{\calorie}{cal}

\newif\ifhighlight
\highlightfalse   

\newcommand{\hl}[1]{%
  \ifhighlight
    \textcolor{blue}{#1}%
  \else
    #1%
  \fi
}

\begin{document}

\title{Simultaneous Learning of Static and Dynamic Charges}

\author{Philipp Stärk}
\affiliation{Stuttgart Center for Simulation Science (SC SimTech),
  University of Stuttgart, 70569 Stuttgart, Germany}
\affiliation{Institute for Computational Physics, University of Stuttgart,
   70569 Stuttgart, Germany}
\author{Henrik Stooß}
\affiliation{Institute for Physics of Functional Materials, Hamburg University of Technology, 21073 Hamburg, Germany}
\author{Marcel F. Langer}
\affiliation{Laboratory of Computational Science and Modeling, IMX, École
  Polytechnique Fédérale de Lausanne,1015 Lausanne, Switzerland}
\author{Egor Rumiantsev}
\affiliation{Laboratory of Computational Science and Modeling, IMX, École
  Polytechnique Fédérale de Lausanne,1015 Lausanne, Switzerland}
\author{Alexander Schlaich}
\email{alexander.schlaich@tuhh.de}
\thanks{Authors contributed equally to this work.}
\affiliation{Institute for Physics of Functional Materials, Hamburg University of Technology, 21073 Hamburg, Germany}
\author{Michele Ceriotti}
\email{michele.ceriotti@epfl.ch}
\thanks{Authors contributed equally to this work.}
\affiliation{Laboratory of Computational Science and Modeling, IMX, École
  Polytechnique Fédérale de Lausanne,1015 Lausanne, Switzerland}
\author{Philip Loche}
\email{philip.loche@epfl.ch}
\thanks{Authors contributed equally to this work.}
\affiliation{Laboratory of Computational Science and Modeling, IMX, École
  Polytechnique Fédérale de Lausanne,1015 Lausanne, Switzerland}

\begin{abstract}
  \input{abstract.tex}
\end{abstract}


\newcommand{\beginsupplement}{%
        \setcounter{table}{0}
        \renewcommand{\thetable}{S\arabic{table}}%
        \setcounter{figure}{0}
        \renewcommand{\thefigure}{S\arabic{figure}}%
        \setcounter{equation}{0}
        \renewcommand{\theequation}{S\arabic{equation}}
}

\newcommand{\R}{\vec{r}}
\newcommand{\pol}{\vec{P}}
\newcommand{\cell}{\vec{c}}


\maketitle

\input{document.tex}

\clearpage
\onecolumngrid


\setcounter{section}{0}
\setcounter{figure}{0}
\setcounter{table}{0}
\setcounter{equation}{0}

\renewcommand{\thesection}{S\arabic{section}}
\renewcommand{\thefigure}{S\arabic{figure}}
\renewcommand{\thetable}{S\arabic{table}}
\renewcommand{\theequation}{S\arabic{equation}}

\begin{center}
    \Large\textbf{Supplementary Information}
\end{center}

\vspace{1cm}

\input{si_content}

\end{document}

%% file: abstract.tex
Long-range interactions and electric response are essential for accurate modeling of condensed-phase systems, but capturing them efficiently remains a challenge for atomistic machine learning.
Traditionally, these two phenomena can be represented by static charges, that \hl{underlie} Coulomb interactions between atoms,
and dynamic charges such as atomic polar tensors---aka Born effective charges---describing the response to an external electric field.
We critically compare different approaches to learn both types of charges within a single model architecture, taking bulk water and water clusters as paradigmatic examples: (1) Learning them independently; (2) Coupling static and dynamic charges based on their physical relationship with a single global coupling constant to account for dielectric screening; (3) Coupled learning with a local, environment-dependent screening factor.
In the coupled case, correcting for dielectric screening is essential, yet the common assumption of homogeneous, isotropic screening breaks down in heterogeneous systems such as water clusters.
A learned, environment-dependent screening restores high accuracy for the dynamical charges.
However, the accuracy gain over independent dynamic predictions is negligible, while the computational cost increases compared to using separate models for static and dynamical charges. This suggests that, despite the formal connection between the two charge types, modeling them independently is the more practical choice for both condensed-phase and isolated cluster systems.

%% file: document.tex
\section{Introduction}
Electric fields influence the structure, dynamics, and reactivity of molecular and condensed-phase systems. Processes such as catalysis, charge transport, or characterization techniques like infrared (IR) spectroscopy depend sensitively on how atoms respond to internal and external electric fields \cite{pan22a,ciampi18a,salles20a}. First-principles electronic-structure methods capture these responses with high accuracy, but their computational cost limits their use for large, heterogeneous systems or long simulation times. Machine-learned interatomic potentials (MLIPs) have begun to address this challenge by providing near-quantum-chemical accuracy for the structure and dynamics of molecules and condensed phases at drastically reduced cost\cite{unke_machine_2021}.
Modern architectures—often graph neural networks with equivariant message-passing—effectively model complex many-atom correlations, albeit within a finite interaction range and model capacity\cite{schutt2017schnet, thomas2018tensor, batzner20223, batatia2022mace}.
However, most MLIPs remain restricted to field-free simulations: they do not natively incorporate the effects of external fields or predict electrical response properties, limiting their applicability in technologically relevant scenarios.

Several strategies have been developed to endow MLIPs with electric-field awareness. One class of methods learns dipoles or local dipole contributions\cite{li26a,kapil24a,falletta24a,stocco25a,veit20a,martin25a}, but these approaches face fundamental challenges:
For ionic (i.e.\ systems with free charges) and periodic systems in general, dipoles are defined only modulo a polarization quantum \cite{resta07a,spaldin12a,stocco25a}, which requires care during training, as the loss function must account for this multi-valuedness.
Other approaches bypass these issues by learning scalar effective charges \cite{dufils23a,li25a} or tensorial quantities such as Born effective charges (BECs), also called atomic polar tensors (APTs) \cite{schienbein23a,joll24a,bergmann_machine_2025}. These models couple external fields to learned charges, generating auxiliary field-dependent forces and thus avoid the conceptual ambiguities of polarization-based methods.
BEC-based models perform well in capturing the linear electric response.
However, these models typically neglect an important physical ingredient: the external electric-field response of atoms is intimately linked to the electrostatic interaction between atoms. Effective charges used to model long-range Coulomb forces reflect internally screened interactions, whereas BECs describe the unscreened response of the electron density to an external field.
Crucially, the distinction between internal (screened) and external (unscreened) responses becomes pronounced in heterogeneous systems such as clusters, interfaces, or molecular mixtures, where assuming a single homogeneous screening value may not be sufficient.

Besides electric-field response, long-range electrostatics have already been incorporated in MLIPs using fixed charges \cite{bartok10a,deng19a}, ML-fitted surrogate charges such as Hirshfeld charges \cite{artrith11a,morawietz12a}, dipole-matching schemes \cite{yao18a}, or global charge-equilibration networks \cite{ko21a,dufils23a,li25a,gao24a}.
Recently, differentiable Ewald summation frameworks \cite{cheng25a,loche_fast_2025} have enabled end-to-end charge learning, removing ambiguities in charge assignment \cite{gross02a,reed85a,hirshfeld77a}.
A few recent studies attempt to combine long-range electrostatics with electric-field response, for example by deriving BECs from end-to-end learned internal charges \cite{zhong25a,kim25a}.
These approaches generally assume homogeneous isotropic screening, which can provide decent qualitative behaviour in bulk systems but may fail in heterogeneous environments, limiting their applicability for general-purpose ML potentials.

This breakdown raises a central open question: should static (effective) and dynamic (BEC) charges be learned together, constrained by their physical relationship, or treated as independent quantities?
Here, we address this question by systematically evaluating both coupled and uncoupled strategies in long-range ML architectures, including cases with environment-dependent screening, to clarify where and why homogeneous screening assumptions break down and how accuracy can be restored.
\hl{We critically compare different models in terms of accuracy, interpretability, and computational cost, using high-quality finite-field reference data for bulk water as a prototypical polar liquid with well-characterized experimental reference data including IR spectra\cite{carlson20a} and water clusters. Water clusters are for example of direct scientific relevance as precursors of atmospheric ice-nucleating particles\cite{knopf_role_2018} and as models for aerosol droplets in the context of airborne pathogen transmission\cite{netz_mechanisms_2020}.}

\section{Theory and Models}

\subsection{Static and Dynamic Charges}

A key concept for describing the response of atomistic systems to externally applied fields is the BEC (or APT) of an atom, which is defined via\cite{schienbein23a,joll24a}
\begin{equation}
  \label{eq:apt_def}
  Z_{i}^{\alpha\beta} = V \pdv{\mathcal{P}^\alpha}{r_i^{\beta}}
  = \pdv{F_{i}^{\beta}}{E^\alpha},
\end{equation}
where $\mathcal{P}^\alpha$ is the $\alpha$-component of the system polarization density, $r_i^\beta$ the $\beta$-component of the position of atom $i$, $V$ is the system volume, $F_i^\beta$ the force on atom $i$ and $E^\alpha$ an externally applied field.
The variables $\alpha, \beta \in \{x, y, z\}$ specify Cartesian components of the tensorial quantities.
The right-hand side of \cref{eq:apt_def} motivates the use of these charges for describing electrostatic interactions with external fields, as it represents the corresponding linear-response coefficient---the change in force induced by an applied external field.

We consider for the moment non-periodic systems, where we define partial charges $q_i^{\text{IR}}$ that reproduce the polarization $\vec{P} = \mathcal{P} V$,
\begin{equation}
  \label{eq:dipole_ir_charges}
  \vec{P} = \sum_j q_{j}^{\text{IR}} \vec{r}_j.
\end{equation}
This allows us to relate BECs $Z_i$ to static charges $q_i^{\text{IR}}$ through\cite{milani10a}
\begin{equation}
  \label{eq:apt_dynamic_static}
  Z_{i}^{\alpha \beta} = q_i^{\text{IR}}\delta^{\alpha \beta} + \sum_j \pdv{q_{j}^{\text{IR}}}{r_i^\beta} r_j^\alpha,
\end{equation}
where the first term mimics the \enquote{static charges} $q_i^{\text{IR}}$, which are sometimes also referred to as the \enquote{IR charge} (as it reproduces the system's dipole and thus yields the infrared spectrum, as we will discuss below).
The second term in \cref{eq:apt_dynamic_static}  describes the charge flux: the change in the dipole moment due to a collective alteration of the charge distribution as the $i$-th atom is displaced by an infinitesimal amount.
Because $Z_i$ encodes these charge redistribution effects, BECs are also called \enquote{dynamic charges}.

\subsection{Long-Ranged Machine Learned Potentials Via Learned Static Charges}

As briefly explained in the introduction, a wide range of methods have been proposed for assigning partial \enquote{static} charges to nuclei from the continuous electron density.
Popular schemes include Mulliken charges\,\cite{mulliken55a}, Hirshfeld charges\,\cite{hirshfeld77a} and the aforementioned IR charges\,\cite{milani10a}.
Each of these methods corresponds to a different definition of the partial charges, and there is no unequivocally ``best'' choice. 
Given this ambiguity, its perhaps not too surprising that many different schemes have been used with similar success to capture long-ranged electrostatic interactions in machine-learned interatomic potentials, including the explicit learning of atomic partial charges\cite{morawietz12a,artrith11a}, the learning of the position of Wannier centers\cite{zhang_deep_2022} and charge equilibration schemes that allow for self-consistent redistribution of atomic charges\cite{ko21a}.
Atomic charges can also be treated simply as fitting parameters to reproduce quantum mechanical energies and forces. This approach is common in classical forcefields\cite{wang_building_2014}. It is also used implicitly by schemes such as LODE, where long-range descriptors have fitting coefficients that correspond to charge multipoles\cite{huguenin-dumittan_physics-inspired_2023}. Recently, the latent Ewald summation framework has popularized this strategy for end-to-end charge learning\cite{cheng25a}.
Instead of targeting a specific charge partitioning scheme, we focus here on the concept of \emph{learned pseudo charges} $q_i$. These are not fitted to any particular definition of atomic charges. Learning static charges indirectly avoids the need to choose an arbitrary partitioning scheme. The model is thus free to infer per-atom quantities $q_i(\xi_i)$ that minimize errors in predicted energies, forces, and, where applicable, Born effective charges (BECs).
The model uses descriptors $\xi_i$ of the local environment of each atom $i$ within a finite cutoff [see \cref{fig:schema} (A,B)]. These descriptors are used to predict both the short-range part of the potential $U_i^{\text{sr}}$ and the atomic pseudo-charges $q_i$. The pseudo-charges are then used to compute the electrostatic potential and thus ultimately the long-range part of the MLIP.
One can go even further and use more general functional forms that do not allow for a transparent interpretation of $q_i$ as atomic static charges.
This approach, briefly introduced in the following section, has been shown to further improve the accuracy of energies and forces\cite{rumiantsev26b}.

\subsection{Uncoupled Static/Dynamic Charge Prediction}

\begin{figure*}
  \includegraphics[width=0.99\textwidth]{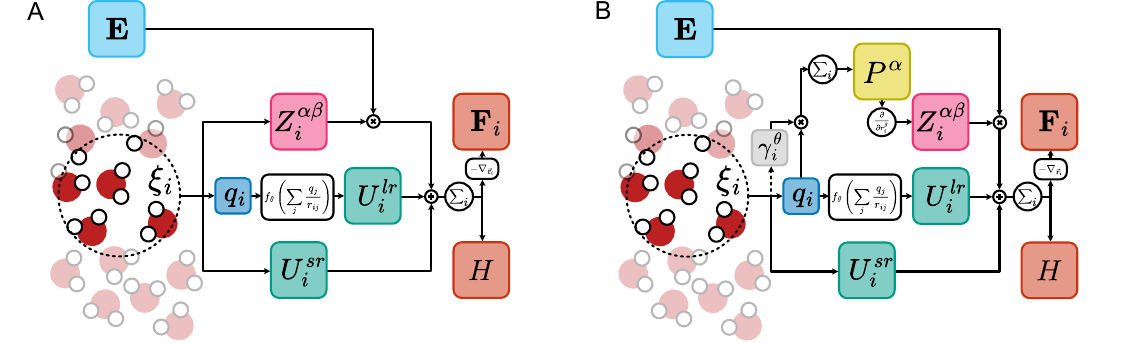}
  \caption{\label{fig:schema}
    Schematic illustration of the two different model architectures.
    A shows the architecture of an \enquote{uncoupled model}, where BECs $Z$ and static charges $q$ are treated independent of one another.
    B shows a coupled architecture where the learned pseudo charges $q_i$ are related to the BECs through the polarization $\vec{P}$, with either a global $\gamma^\theta_i = \text{const.}$ or local screening parameters $\gamma^\theta_i$.
    The final output of each model is the total force per atom $\vec{F}_i$ and the electrostatic enthalpy $H$ (see \cref{eq:ext_field_energy}), which reduces to the internal energy $H = U$ in the case of no external fields, $\vec{E} = 0$.
  }
\end{figure*}

In the present work, the long-ranged contribution to the potential is constructed using learned pseudo-charges $q_i$ introduced above.
Without PBC, the electrostatic potential
\begin{equation}
  \label{eq:coulomb}
  V_i = \sum_{j\neq i} \frac{q_i}{4\pi \epsilon_0\, r_{ij}}\,,
\end{equation}
is computed through a algorithmic differentiable (AD) Coulomb solver. Periodic image contributions are handled via an AD implementation of the Ewald summation\cite{loche_fast_2025}.
To integrate the interactions associated with the Coulomb potential into an energy-decomposable MLIP, we express the long-range energy as
\begin{equation}
    \label{eq:physical_lr_en}
  U^{\mathrm{lr}} = \sum_i V_i (\{q_j\}_j)\, q_i(\xi_i).
\end{equation}

Beyond this physical definition of $U^{\mathrm{lr}}$, we also consider a more flexible long-range formulation introduced in Ref.~\citenum{rumiantsev26b}.
Here, the electrostatic potential is processed by a learnable function $f_\theta$, yielding
\begin{equation}
\label{eq:lr_energy_lorem}
  U^{\mathrm{lr}} = \sum_i f_\theta(V_i (\{q_j\}_j, \xi_i)\,.
\end{equation}
This modification, referred to as \enquote{Learning Long-Range Representations with Equivariant Messages} (LOREM) has been shown to significantly reduce errors across multiple prediction tasks\cite{rumiantsev26b}.
We note that in this work we restrict the architectures to learned scalar pseudo charges only, as opposed to the more general equivariant tensorial pseudo charges proposed in Ref.~\citenum{rumiantsev26b}.
When the learnable function is set to be the identity, $f_\theta = \mathrm{id}$, the model reduces to \cref{eq:physical_lr_en} and retains interpretability of the pseudo charges. We refer to this latter version as the \emph{physical} long-range architecture.

The simplest extension of long-ranged MLIPs to also predict BECs is to treat them as independent outputs without connection to the static pseudo charges.
An architecture capable of this is shown in \cref{fig:schema}A.
Here, the pseudo charges $q_i(\xi_i)$ are predicted solely for computing $U^{\mathrm{lr}}$, while the BECs constitute a second, independent prediction target derived from the same local environment representation $\xi_i$.
This design allows the model to learn static and dynamic charge responses separately, without enforcing any explicit relation between them.

Because we employ an equivariant descriptor $\xi_i$, predicting $Z_i$ becomes straightforward.
To obtain the Born effective charge tensor $Z_i^{\alpha\beta}$ from the local atomic environment, we map the equivariant spherical-harmonic features of each atom through a small neural module.
The descriptor (restricted to features up to two angular channels) is first processed by several dense layers applied to the spherical channels.
An equivariant tensor-coupling step based on Clebsch-Gordan coefficients mixes the angular components to form all symmetry-allowed rank-2 contributions, which are subsequently linearly recombined into a $3\times 3$ matrix, which is fitted to BEC labels derived from DFT (see methods \cref{sec:meth-dataset})\cite{unke24a}.
This construction ensures that the predicted tensors transform correctly under rotations.
Such an approach is essentially the extension of previous works\cite{schienbein23a,schmiedmayer24a} to long-ranged architectures.

\subsection{Coupled Static/Dynamic Charge Prediction}
\label{ssec:coupled_theory}
A second possible strategy is to attempt to approximate the infrared charges $q_i^{\text{IR}}$ via the learned pseudo charges $q_i(\xi_i)$ to construct $Z_i$ via \cref{eq:apt_dynamic_static,eq:apt_def}, as has been proposed in recent works\cite{zhong25a,kim25a,schmiedmayer26a}.
However, when $q_i(\xi_i)$ are learned through minimization of energy and force errors using \cref{eq:coulomb,eq:physical_lr_en}, these quantities already include screening effects from the instantaneously responding electronic background.
For homogeneous bulk systems\cite{zhong25a} or isolated molecules in vacuum\cite{kim25a}, the screening can be approximated as homogeneous and isotropic. This leads to a simple proportionality between learned and infrared charges
\begin{equation}
  \label{eq:gamma_scaling}
  q_i^{\mathrm{IR}} = \gamma\, q_i\hl{(\xi_i)}\,.
\end{equation}
Here, the coupling parameter $\gamma$ can be interpreted in terms of the high-frequency dielectric permittivity, $\gamma \approx \sqrt{\epsilon_\infty}$\cite{zhong25a,leontyev09a,leontyev10a}.
The underlying assumption of the latter relation is that learned pseudo charges correspond to physically interpretable partial charges and has been shown to allow for a-posteriori prediction of $Z_i$ in a large variety of systems, if one assumes $\epsilon_\infty$ between 1 and 1.83\cite{zhong25a,kim25a}.
Here, we will refrain from interpreting the screening parameter as directly related to $\epsilon_\infty$. Instead, we simply treat $\gamma$ as a learnable parameter that scales the learned pseudo charges to infrared charges. This scaling is determined by requiring that the resulting charges reproduce BEC targets obtained from DFT calculations.
However, as we will show below, the assumptions of homogeneous and isotropic screening break down in inhomogeneous environments such as interfaces, where the dielectric response is well known to become both anisotropic and spatially varying\cite{kornyshev86a,stern03a,ballenegger05a,stark25d}. A simple way to resolve the homogeneity assumption is to introduce a \emph{local} screening factor,
\begin{equation}
    \label{eq:local_screen}
  q_i^{\text{IR}} = \gamma_i(\xi_i)\, q_i(\xi_i),
\end{equation}
where the scalar coefficient $\gamma_i$ is predicted from the local environment $\xi_i$, which follows smoothly from \cref{eq:gamma_scaling}.
This local formulation captures spatial inhomogeneities, but it still enforces \emph{isotropic} screening.
Promoting $\gamma_i$ to a \emph{tensorial}, environment-dependent quantity would be equivalent to directly predicting the full dynamic charge $Z_i$.
Consequently, uncoupled learning of $Z_i$ captures both anisotropy and inhomogeneity, whereas scalar local screening only accounts for the latter.

Using the proposed coupling relations \cref{eq:gamma_scaling,eq:local_screen} to predict $Z_i$ via \cref{eq:apt_dynamic_static} is the basis for the second family of MLIPs that we study, the \emph{coupled models}.
These models, shown in \cref{fig:schema}B, use the predicted $q_i(\xi_i)$ to construct the BECs.
We explore two variants of these coupled models.
(i) In the coupled, global approach, the MLIP is first trained only on energies and forces. Afterwards, a single global screening parameter $\gamma$ (see \cref{eq:gamma_scaling}) is fitted a-posteriori to minimize the BEC error on the validation set.
We intend this strategy to closely follow the strategy of Refs.~\citenum{zhong25a,kim25a}.
(ii) In the coupled, local approach, we drop the assumption of a uniform screening factor.
Instead, the model predicts a local, environment-dependent $\gamma_i$ (see \cref{eq:local_screen}), learned jointly with energies, forces, and BEC labels.
This allows us to test whether relaxing the homogeneity assumption improves BEC prediction accuracy, especially in inhomogeneous systems.

For all coupled architectures an issue arises when incorporating periodic boundary conditions.
A naive definition of the polarization density via \cref{eq:dipole_ir_charges,eq:gamma_scaling} can result in ill-defined values, due to the periodic boundaries.
This problem can be mitigated in molecular systems by learning molecular dipole contributions~\cite{bereau15a}, but this approach is incompatible with the atom-centered, point-charge picture adopted by the models we consider in the present work.
\citet{zhong25a} solved this by utilizing an artificial complex phase.
We instead calculate $Z^i$ from the learned pseudo charges via their positions by
\begin{equation}
  \label{eq:apt_mic}
  \pdv{P^\alpha}{r_i^\beta} = q_i^{\text{IR}} \delta^{\alpha \beta} + \sum_j \pdv{q_j^{\text{IR}}}{r_i^\beta} \left(r_j^\alpha - r_i^\alpha\right)_{\text{PBC}}\,,
\end{equation}
where the index \enquote{PBC} accounts for periodic boundary conditions when computing the distances.
We explain the motivation behind \cref{eq:apt_mic} in detail in section I of the supporting information.
Equation \ref{eq:apt_mic} directly follows from recognizing that for neutral systems $\vec{P}$ is translationally invariant and can thus always be formulated with respect to particle distances, which removes complications due to ambiguous positions in systems with toroidal boundary conditions.
For computational efficiency, it is necessary to reformulate \cref{eq:apt_mic}, which greatly reduces the computational cost of inference and training, following a previously reported approach for computing heat fluxes \cite{langer23a,langer23b}.
This optimized implementation is presented in detail in section II of the SI. 

\section{Results and discussion}

\begin{figure}
  \includegraphics{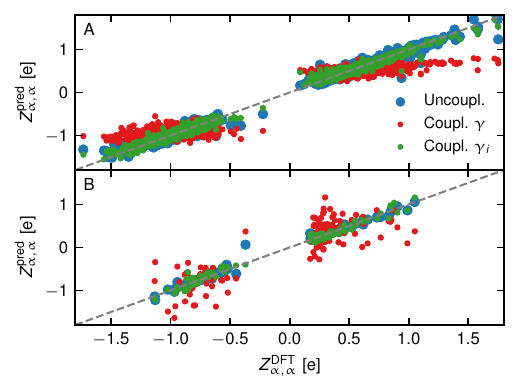}
  \caption{\label{fig:scatter_plots_apt}
  Scatter plot of diagonal components of the BECs $Z^{\alpha\alpha}$ comparing DFT values with model predictions for bulk water (A) and water clusters (B).}
\end{figure}

\Cref{fig:scatter_plots_apt} compares parity plots of the diagonal BEC components $Z^{\alpha \alpha}$ for bulk water (panel A) and water clusters (panel B) using the uncoupled model and two coupled variants. Unless noted otherwise, all models in this section are the \emph{physical models} and were trained on a dataset combining both bulk and cluster configurations (LOREM-model results are reported in Figure S2 of the supporting information). See \cref{sec:meth-training} for full training details.

Overall, the uncoupled and the coupled model with local screening ($\gamma_i$) both reproduce the BEC labels accurately for the bulk and for the clusters. The global-screening model, however, shows a clear deterioration in performance on both subsets. This stems from the global model’s homogeneous-screening assumption: as expected from our discussion above, a single $\gamma$ cannot simultaneously capture the different effective dielectric responses present in homogeneous bulk and inhomogeneous cluster environments.
For this global-$\gamma$ model---where $\gamma$ was obtained by an a-posteriori fit to the BEC labels of the validation-set---we find $\gamma=1.974$.
While one could reduce the error further by fitting distinct screening factors for bulk and cluster subsets, this strategy requires identifying and splitting the dataset into several classes based on their structure. As dataset size and diversity grow, such manual classification becomes increasingly impractical.

\begin{figure}
  \includegraphics{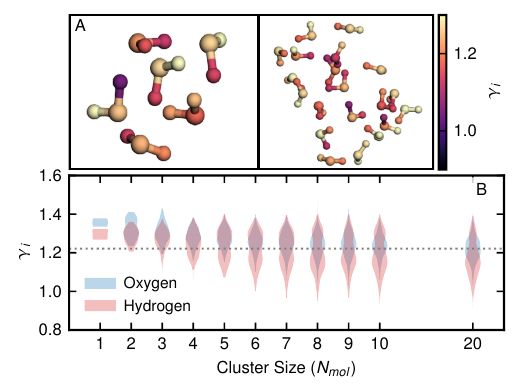}
  \caption{\label{fig:screening_plot}
    A:~Snapshot of water clusters with 6 and 20 molecules. Atoms are colored according to their local screening values $\gamma_i$. An interactive view is provided online \url{https://doi.org/10.24435/materialscloud:fs-8h}.
    B:~Distribution of the local screening values from the coupled $\gamma_i$ model as a function of water cluster sizes. Gray dashed line shows the average value of $\gamma_i$ for bulk structures.
  }
\end{figure}

The local screening varies strongly in clusters. \Cref{fig:screening_plot} shows the distributions of local screening values $\gamma_i$ predicted by the coupled local-$\gamma_i$ model as a function of the number of molecules $N_{\text{mol}}$ per cluster.
For small clusters the predicted $\gamma_i$ values are tightly distributed, whereas the distributions broaden as cluster size increases.
Notably, we observe pronounced differences for dangling OH groups, highlighted by the color-coded structures in \cref{fig:screening_plot}A.
This trend is also reflected in the bimodal distribution of hydrogen-atom $\gamma_i$ values in cluster structures, which we attribute to the distinction between hydrogens engaged in hydrogen bonding and those exposed at the cluster surface.

In \cref{fig:screening_plot}B, the dashed gray line indicates the mean screening value for bulk structures, $\overline{\gamma_i} = 1.22 \pm 0.09$, which the $\gamma_i$ values approach for larger clusters, as expected.
Together, these findings provide a physically interpretable explanation for the coupling between learned pseudo charges, infrared charges, and BECs.
Accurate modeling in heterogeneous environments requires this coupling to depend on the local atomic environment. By contrast, a single global screening parameter fails to capture the variability present in inhomogeneous structures.

\begin{figure}
  \includegraphics{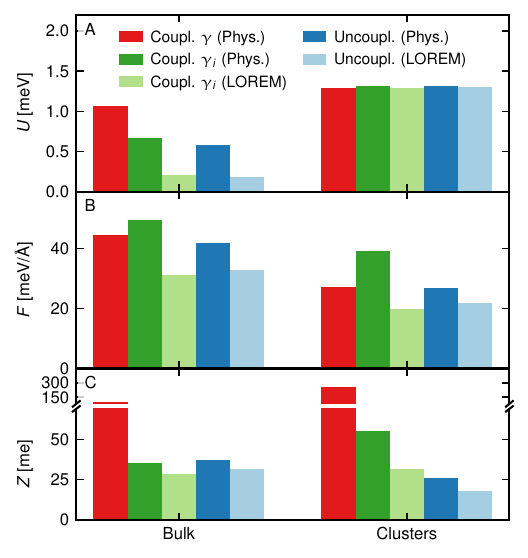}
  \caption{\label{fig:error_bars}
  Mean average errors (MAE) on the validation sets for the energy $U$ (A), forces $F$ (B) and the BEC diagonal elements $Z$ (C). Different colors depict different models. Light color bars show physical models while desaturated bars show the LOREM models.}
\end{figure}

In order to estimate the general fidelity of the different MLIP architectures, we compare the different models in terms of their mean absolute errors (MAEs) on validation sets for energies, forces and BECs (\cref{fig:error_bars}).
Discussing first the physical models, we find differences in energy and force errors to be minor, with the uncoupled model performing overall the best.
As already observed in \cref{fig:scatter_plots_apt}, the coupled global $\gamma$ model performs significantly worse on BEC predictions, while the uncoupled and coupled local $\gamma_i$ models achieve similar accuracies for predicting BECs, underscoring the need for environment dependent coupling between learned pseudo charges and BECs for these systems.
Importantly, modeling the coupling increases the computational cost by a constant prefactor relative to the uncoupled model, while the asymptotic scaling with system size remains unchanged (see Fig. S1 in the Supporting Information).

So far, all models were trained utilizing the physically motivated Coulomb interaction of \cref{eq:physical_lr_en}.
To investigate if a more expressive but less physically interpretable model utilizing \cref{eq:lr_energy_lorem} may perform even better, we also trained models with the LOREM formulation.
As can be seen in \cref{fig:error_bars}, this modification improves performance on all targets with a particularly strong improvement in bulk potential energy errors.
The parity plot (similar to \cref{fig:scatter_plots_apt}) for the LOREM models are presented in the supporting information (Figure S2), showing qualitative agreement with the physical models and mirroring the need for a location dependent coupling parameter to achieve good BEC accuracy on cluster and bulk structures.
However, as shown in Figure S3, the local screening values $\gamma_i$ predicted by the LOREM models are no longer physically interpretable due to the learnable mapping $f_\theta$.

\begin{figure}
  \includegraphics{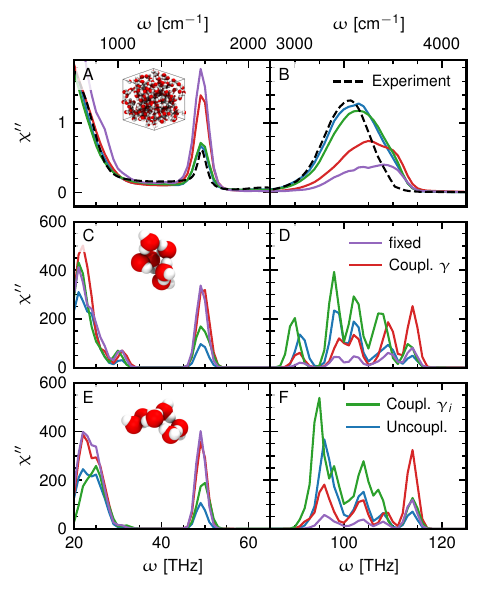}
  \caption{\label{fig:dielectric_spectrum}
  Imaginary part of the susceptibility spectrum of periodic bulk water at $T = \SI{300}{\kelvin}$ \hl{(A+B)}, cage \hl{(C+D)} and book \hl{(E+F)} configuration of water hexamer clusters at $T = \SI{10}{\kelvin}$. \hl{We show results for the three model architectures discussed in this work and for reference the result of a simple constant scalar charge analysis (purple line).} The black dashed line shows experimental bulk reference spectrum taken from Ref.~\citenum{carlson20a}.}
\end{figure}

Finally, we assess the behavior of the different physical models in predicting infrared spectra from molecular dynamics simulations. Computational details are provided in method \cref{sec:meth-IR}. \Cref{fig:dielectric_spectrum} shows the imaginary part of the complex susceptibility for periodic bulk water at $T = \SI{300}{\kelvin}$ and for the cage and book configurations of water hexamer clusters at $T = \SI{10}{\kelvin}$. \hl{The spectra are organized into two columns: the left column (panels A, C, E) shows the OH-bending mode around $\sim \SI{50}{\tera\hertz}$, while the right column (panels B, D, F) shows the OH-stretching mode around $\sim \SI{100}{\tera\hertz}$.}
We note that nuclear quantum effects are neglected, so comparison with experiment is qualitative, although error cancellation in GGA functionals yields reasonable agreement for bulk water.
\hl{To contextualize the sensitivity of spectra dynamics in the prediction of BECs, we also provide results for an analysis based on constant scalar charges.
To this end, we calculate the average diagonal component of the BECs per element and assign this charge to all O and H atoms for the analysis of the entire trajectory. The results of this procedure are given by the purple lines in \cref{fig:dielectric_spectrum}.}

\hl{Focusing first on a comparison of models in the bending mode in panels A, C, E, a consistent hierarchy emerges: the fixed charge model (based on atom-type averaged charges), global screening ($\gamma$), local screening ($\gamma_i$), and uncoupled models estimate the bending mode amplitude in decreasing order.}
\hl{In bulk water, where we also provide an experimental reference, this ordering is particularly evident. The global $\gamma$ model strongly overestimates the bending intensity (A) and underestimates the stretching band (B), which is additionally shifted to higher frequencies.
The fixed-charge model shows similar but slightly more pronounced deviations from the experimental reference.}

\hl{In contrast, both the $\gamma_i$ and uncoupled models reproduce intensities and peak positions much more accurately, in closer agreement with experiment.}
\hl{Discussing in more detail the results for the water hexamers, we find similar but slightly less pronounced trends.
As the spectral features of these isomers have been discussed extensively elsewhere \cite{carlson20a,perez12a,wang_ir_2013}, we focus here on differences between models.
Overall, all models reproduce the main spectral features, indicating consistent nuclear dynamics. Differences arise mainly in intensities and subtle frequency shifts.
The bending-mode intensity again increases with decreasing model flexibility (C, E), following the same hierarchy as discussed above, while differences in the stretching region (D, F) are more subtle.}
The LOREM models give spectra \hl{supporting the same conclusion}, as shown in Figure S4 in the Supplemental Information.
\hl{In summary, these results are consistent with the model errors discussed above, in that the global screening model shows the largest deviations in predicted infrared spectra, including for the experimentally accessible bulk water case.
This reinforces the conclusion that a more local treatment of the coupling between latent and dynamic charges, as realized in the $\gamma_i$ model, or the complete decoupling in the uncoupled model, more accurately captures the charge redistribution underlying IR activity, particularly for the OH bending/stretching part of the spectrum, where the dynamic part of BEC contributions are expected to be pronounced.
At the same time, the overall spectral shape is already reasonably well reproduced even with the simplified fixed-charge analysis, indicating that IR spectra are primarily governed by the nuclear dynamics and the symmetry of vibrational modes, while still reflecting the improved physical consistency of more flexible charge-coupling schemes.}

\section{Conclusions}
Machine-learned interatomic potentials increasingly attempt to incorporate long-range electrostatics and external-field coupling by introducing atom-centered static charges and, in some cases, by exploiting \hl{a physically motivated} relationship between static and dynamic charges.
However, an explicit assessment of whether this physical motivation \hl{enables reliable prediction of} dynamical charges without explicitly training on them is lacking, \hl{and there is a lack of systematic studies evaluating whether coupled architectures with an explicit dynamic-charge target outperform approaches that treat static and dynamic charges separately.}

In this work, we systematically \hl{examine} this question within models that incorporate explicit electrostatic interactions. Using water clusters and bulk water as controlled test systems—where high-quality reference data are available and long-range polarization effects are essential—we demonstrate that even in this comparatively simple setting, static charges learned as coefficients of a Coulomb term are only \hl{correlated} with Born effective charges (BECs). \hl{As we show, the relationship is not captured well with a single factor across heterogeneous environments, breaking down when fitting on systems such as clusters of varying size.}
We show that restoring quantitative agreement within coupled models for heterogeneous datasets requires employing spatially varying screening coefficients. While this improves accuracy, it removes the anticipated simplicity of the physically constrained approach and \hl{does not provide a clear computational advantage} over directly learning BECs as an independent target. In contrast, treating dynamic charges as a separate, well-defined observable yields higher accuracy, improved infrared intensities, and lower complexity.

Even though it is tempting to avoid the explicit calculation of reference \hl{BECs}, especially given that observables such as IR spectra are \hl{comparatively insensitive to moderate errors} in the predicted dynamical charges, \hl{our results show that measurable and systematic differences nevertheless emerge, in particular for bulk water where experimental reference data are available. In this case, models incorporating local screening or decoupled charge responses reproduce the relative intensities and peak positions more accurately than approaches based on global screening or fixed charges.}
Our results indicate that \hl{whenever heterogeneous datasets are considered, quantitative accuracy is strongly improved by explicit training on BECs.}
\hl{For these reasons}, we recommend (i) using electrostatic potentials to capture long-range interactions without attributing physical meaning to static charges, which are \hl{inherently model-dependent quantities}, and (ii) learning dynamic charges explicitly and independently from well-defined \emph{ab initio} BEC data.

\section{Methods}

\subsection{Dataset construction}\label{sec:meth-dataset}

The bulk water structures are taken from Ref.~\citenum{cheng19a}.
The water cluster structures are composed from three sources: (1) the water cluster subset of Hobza’s benchmark energy and geometry data base (BEGDB)\cite{temelso11a}, (2) the WATER27 component of the GMTKN24 and GMTKN30 benchmark suites\cite{bryantsev09a} and (3) dimer and trimer structures from Ref.~\citenum{nguyen18a}. These sets contain very distinct geometrical structures from 1 to 20 water molecules. To extend the dataset size we performed molecular dynamics simulations for each cluster size at a temperature of $400\,\mathrm{K}$ for $500\,\mathrm{ps}$ with a $0.5\,\mathrm{fs}$ timestep
using the universal PET-MAD model\cite{mazitov25a}.
We combined the initial structures with the MD runs and selected 2000 structures using farthest point sampling using the PET-MAD descriptor\cite{mazitov25b}.

For consistent labeling of the structures, we recalculated energies, forces and BECs for all structures with DFT at the revPBE-D3 level using the CP2K package\cite{kuhne20a}.
These calculations employed the revised Perdew--Burke--Ernzerhof (PBE) exchange--correlation functional, a DZVP-MOLOPT atom-centered basis set, and Goedecker--Teter--Hutter (GTH) pseudopotentials.

To compute BECs via finite electric fields we start from the right-hand-side of \cref{eq:apt_def}, indicating that BECs can be obtained as the derivative of atomic forces with respect to externally applied fields.
This relation, strictly only true in the infinitesimal limit, can be approximated via central finite differences, giving us
\begin{equation}
  \label{eq:finite_diff_apt}
  Z_i^{\alpha \beta} \approx \frac{F_i^\alpha(E^\beta/2) - F_i^\alpha(-E^\beta/2)}{E^\beta},
\end{equation}
where $E^\beta$ is the magnitudes of the externally applied fields in Cartesian direction $\beta$ and $F_i^\alpha$ is the $\alpha$ Cartesian component of the force on atom $i$.
This reduces the calculation of the per-atom BECs to six plus one single-point calculations in DFT, where reusing the initial wave-function guess further significantly reduces the computational cost.
We show in Figure~S5 of the SI, that the estimated values for the BECs from this scheme are stable across a very wide range of externally applied field strengths. To apply homogeneous external fields we use the implementation by \citeauthor{souza02a} and \citeauthor{umari02a}\cite{souza02a,umari02a} (see also Ref.~\citenum{stengel09b}).
For production calculations, finite field strengths of
$\SI{0.026}{\volt\per\angstrom}$ were applied for the central finite-difference scheme in \cref{eq:finite_diff_apt}.

\subsection{Training Procedure}\label{sec:meth-training}

We construct the model to output the enthalpy $H$, such that the total force on atom $i$ is given by $\vec{F}_i = -\nabla_{\vec{r}_i} H$ via
\begin{equation}
  \label{eq:ext_field_energy}
  H = \left(\sum_jU_j^{\text{model}} - \vec{r}_j ( Z_j \vec{E} )\right) - \vec{P} _{\text{ref}} \cdot \vec{E}\,,
\end{equation}
where $\vec{P} _{\text{ref}}$ is the dipole of a reference configuration which serves to obtain a consistent definition in periodic settings~\cite{resta07a}.
The internal energy per atom $i$ is simply given by the sum of long and short-ranged contributions $U_i^{\text{model}} = U_i^{\text{sr}} + U_i^{\text{Coul}}$ .
The force given by the gradient of \cref{eq:ext_field_energy} with respect to $\vec{r}_i$ is thus given by:
\begin{equation}
  \label{eq:force}
  F_i^\alpha = \left(-\pdv{r_i^\alpha} \sum_j U_j^{\text{model}}\right) + \sum_\beta Z_i^{\alpha,\beta} E^\beta,
\end{equation}
neglecting second order derivatives of $Z_i$ with respect to $ r_i$.
Provided $\vec{P}$ is known for a reference configuration (e.g.\ via an ab-initio calculation for the starting structure), the enthalpy $H$ is well-defined through $\vec{P} _{\text{ref}}$ during a simulation run.
This procedure is in principle similar to what has been proposed in Ref.~\citenum{falletta24a}, even though here the ambiguity in the definition of $H$ is made explicit.
We note the absence of $\vec{P} _{\text{ref}}$ for the force in \cref{eq:force}, hence forces are always well-defined, even when $\vec{P}$ is ambiguous.

Finally, training of all models is performed by constructing a typical force and energy loss function via an $L_2$ norm, i.e.
\begin{align}
  \label{eq:loss}
  \mathcal{L} = &\alpha_F (\vec{F} _{\text{model}} - \vec{{F}} _{\text{DFT}})^2 
  + \alpha_U (U _{\text{model}} - U _{\text{DFT}})^2 \nonumber\\
   & + \alpha _{\text{BEC}} \sum_i (Z_i^{\mathrm{DFT}}- Z_i^{\text{model}})^2
  + \mathcal{L}_{\text{neutral}}\,,
\end{align}
where $\alpha_F$, $\alpha_U, \alpha _{\text{BEC}}$ are hyper-parameters of the learning procedure.
For some experiments, we set $\alpha _{\text{BEC}}$ of the coupled global $\gamma$ model to zero in order to reproduce the work in Ref.~\citenum{zhong25a}, where BECs are a fitted posteriori  from learned pseudo charges.
To encourage charge-neutral predictions, we include an explicit neutrality penalty in the loss function for all models. This penalty term suppresses artifacts associated with the uniform background charge inherent to Ewald-sum–based electrostatics \cite{barr12a} and is defined as
\begin{equation}
  \label{eq:stability_neutral}
  \mathcal{L} _{\text{neutral}} = \alpha _{\text{neutral}} \left(\frac{\sum_i q_i}{\sum_i \abs{q_i} + \epsilon}\right)^2,
\end{equation}
with $\epsilon = \SI{1e-12}{}$ a small number to ensure numerical stability and $\alpha _{\text{neutral}}$ again a hyper-parameter.
During all training experiments reported in this work all loss weights were set to 1.

For the short-range features, we follow the architecture described in Ref.~\citenum{rumiantsev26b}. All models use a cutoff radius of $r_{\text{c}} = \SI{5.0}{\angstrom}$ and spherical harmonics up to degree 6, with eight spherical channels. The radial dependence is represented using 32 radial basis functions, and the chemical embedding uses eight channels. We employ a cosine cutoff to ensure smooth behavior at the cutoff distance, and perform the radial expansion using basic Bernstein basis functions\cite{lorentz86a}. The network uses the SiLU activation function. No message passing is applied, and for each a single floating-point value is predicted, representing its pseudo charge.

Training is performed using the Adam optimizer with an initial learning rate of $\SI{6e-5}{}$ and a batch size of one structure. Optimization proceeds until the loss no longer improves significantly. For validation, we use a random 80:20 train–validation split.

For the training of the coupled models with global screening parameter $\gamma$, we first train the model purely on energy and force labels.
Utilizing the pseudo charge predictions of this model (through \cref{eq:apt_dynamic_static} for non-periodic structures and \cref{eq:apt_mic} for periodic structures) we determine a single $\gamma$ value which minimizes the mean average error on BEC predictions on the entire validation set, including clusters and bulk structures simultaneously.
This value is then used for all subsequent evaluations and simulations of the model.

\subsection{Molecular Dynamics and IR Spectra}
\label{sec:meth-IR}

All simulations were carried out using the atomic simulation environment (ASE)\cite{hjorthlarsen17a}.
A time step of $\SI{0.5}{\femto\second}$ was used in conjunction with a Bussi-Parinello thermostat\cite{bussi07a} to simulate bulk water and water clusters at constant temperature and constant volume.
For simulations of clusters, we removed the center of mass velocity and rotation and performed ten runs per model and structure which were then averaged over.
\hl{All results for clusters shown in the main work are obtained as the average over 95 independent simulations of $\SI{50}{\pico\second}$ per system and model.
For the bulk systems, twenty simulations of length $\SI{100}{\pico\second}$ are used.
For comparison the Supporting Information shows results for the LOREM models obtained from five independent simulations for each system and model.}
The polarisation time derivative follows from the chain rule\cite{schienbein23a}
\begin{equation}
\label{eq:time_deriv_m}
\dot{\vec{P}} = \sum_i Z_i \vec{v}_i,
\end{equation}
with $\vec{v}_i$ the particle's velocity.
The frequency-dependent complex electric susceptibility is obtained from the polarisation-polarisation time correlation function\cite{carlson20a}
\begin{equation}
\chi(\omega) = -\frac{1}{3 V \varepsilon_0 k_B T}
\int_0^\infty dt\, e^{-2\pi i \omega t}\,
\frac{d}{dt}\langle \vec{P}(0)\cdot\vec{P}(t)\rangle\,,
\end{equation}
where $\vec{P}(t)$ is calculated from \cref{eq:time_deriv_m} via numerical integration.
Using the Wiener--Khinchin theorem, we compute the dissipative imaginary part $\chi''$ from the polarisation spectrum as
\begin{equation}
\label{eq:wiener_khinchin}
\chi''(\omega)
= \frac{\pi \omega}{3 L_t V \varepsilon_0 k_B T}
\left|\tilde{\vec{P}}(\omega)\right|^2,
\end{equation}
where $\vec{\tilde{P}} (\omega)$ is the Fourier transform of the polarisation time series \hl{and where $L_t$ is the length in time of the polarization $\vec{P}(t)$.}

\hl{For the reference analysis method utilizing fixed charges, we use the trajectory obtained from the uncoupled model and take the average $\overline{Z}_{\alpha \alpha}$ of the diagonal BEC components over $\SI{50}{\pico\second}$ per atomic species.
The dipole IR spectra are then calculated via \cref{eq:wiener_khinchin} via $\vec{P}(t) = \sum_i \overline{Z}_{\alpha \alpha} r_i(t) $.}

\section*{Author contributions}
P.L., M.C. and A.S. designed the study.
P.S. implemented the method, trained the models, performed simulations and analysis. P.L. curated the datasets, created the visualizations, and assisted with implementation and simulations.
M.F.L. and E.R. contributed significantly to the design of the models, with E.R. additionally supporting model training and figure design.
H.S. performed the density functional theory calculations.
M.C. and A.S. supervised the project.
All authors contributed to the writing of the manuscript and analysis of the results.

\section*{Conflicts of interest}
There are no conflicts to declare.

\section*{Code availability}
The trained models, training scripts, CP2K input files to generate the dataset, and the code to reproduce this study is available on GitHub at \url{https://github.com/pstaerk/si_charge_learning_bec}.

\section*{Data availability}
The atomic structures used to train the models are available from \url{https://doi.org/10.24435/materialscloud:fs-8h}.
In the same entry we also provide a chemiscope visualization for the validation set of the water clusters for the local gamma models containing the structures, screening values and learned pseudo charges.

\section*{Acknowledgements}
We thank Michelangelo~Domina, Philipp~Schienbein, Bingqing~Cheng, Venkat Kapil and Matthias Kellner for the insightful discussions.
The work of A.S. and P.S. was funded by Deutsche Forschungsgemeinschaft (DFG, German Research Foundation) under Germany's Excellence Strategy - EXC 2075/1 – 390740016 and further supported by EXC 3120/1 – 533771286.
We acknowledge the support by the Stuttgart Center for Simulation Science (SimTech).
H. S. was funded by the Deutsche Forschungsgemeinschaft (DFG, German Research Foundation) under SFB 1333/2 – 358283783.
M.F.L. acknowledges funding from the German Research Foundation (DFG) under project number 544947822. M.C. and P.L. acknowledge funding from the NCCR MARVEL, funded by the Swiss National Science Foundation (SNSF, grant number 182892) and from the European Research Council (ERC) under the European Union’s Horizon 2020 research and innovation programme (grant agreement No 101001890-FIAMMA).

\bibliography{
  bibliography/Literature.bib,
  bibliography/philip.bib,
  bibliography/egor.bib
}

%% file: si_content.tex
\title{Supplemental~Information: Simultaneous Learning of Static and Dynamic Charges}

\author{Philipp Stärk}
\affiliation{Stuttgart Center for Simulation Science (SC SimTech),
  University of Stuttgart, 70569 Stuttgart, Germany}
\affiliation{Institute for Computational Physics, University of Stuttgart,
   70569 Stuttgart, Germany}
\author{Henrik Stooß}
\affiliation{Institute for Physics of Functional Materials, Hamburg University of Technology, 21073 Hamburg, Germany}
\author{Marcel F. Langer}
\affiliation{Laboratory of Computational Science and Modeling, IMX, École
  Polytechnique Fédérale de Lausanne,1015 Lausanne, Switzerland}
\author{Egor Rumiantsev}
\affiliation{Laboratory of Computational Science and Modeling, IMX, École
  Polytechnique Fédérale de Lausanne,1015 Lausanne, Switzerland}
\author{Alexander Schlaich}
\email{alexander.schlaich@tuhh.de}
\thanks{Authors contributed equally to this work.}
\affiliation{Institute for Physics of Functional Materials, Hamburg University of Technology, 21073 Hamburg, Germany}
\author{Michele Ceriotti}
\email{michele.ceriotti@epfl.ch}
\thanks{Authors contributed equally to this work.}
\affiliation{Laboratory of Computational Science and Modeling, IMX, École
  Polytechnique Fédérale de Lausanne,1015 Lausanne, Switzerland}
\author{Philip Loche}
\email{philip.loche@epfl.ch}
\thanks{Authors contributed equally to this work.}
\affiliation{Laboratory of Computational Science and Modeling, IMX, École
  Polytechnique Fédérale de Lausanne,1015 Lausanne, Switzerland}

\maketitle

\beginsupplement
\section{Calculating BECs from Learned Charges in Periodic Systems: In Principle}
In the main text, we indicated that care has to be taken in using infrared charges $q _{\text{IR}}^i$ for the calculation of $Z^{i}_{\alpha\beta}$ in systems with periodic boundary conditions (pbc). The reason for this is that in pbc, particle positions are only defined up to integer multiples $n_a$ of the lattice vectors $\cell_a$. In other words, the arbitrary choice of which unit cell to pick and where to draw its boundary can add or remove offsets from $\R_j$. This changes the value of $\pol$, and, by the chain rule, also for its derivatives $Z^{i}_{\alpha\beta}$ when using the  \enquote{naive definition} of $\vec{P} = \sum_j q_j \vec{r}_j$. Let us consider the case where the boundary is shifted such that only one position, $i$, crosses it:
\begin{equation}
    \R_j \longrightarrow \R_j + \sum_{a=1}^3 n_a \cell_a
\end{equation}
This changes the polarization
\begin{equation}
    \sum_j q_j \vec{r}_j \longrightarrow \sum_j q_j \vec{r}_j + q_i \sum_{a=1}^3 n_a \cell_a 
\end{equation}
and consequently the derivatives
\begin{equation}
    \frac{\partial}{\partial \R_k} \sum_j q_j \vec{r}_j \longrightarrow \frac{\partial}{\partial \R_k} \sum_j q_j \vec{r}_j + \left( \frac{\partial}{\partial \R_k} q_i \right) \sum_{a=1}^3 n_a \cell_a 
\end{equation}
in an arbitrary way that the model cannot learn and leads to noisy predictions observed by \citeauthor{zhong25a}~\cite{zhong25a}.

We propose to fix this ambiguity by transitioning to relative, boundary-invariant, positions for the $\R_j$ used in constructing $\pol$.
To motivate this, we note that since $\sum_j q_j = 0$ due to charge neutrality, an offset can be added to $\R_j$ without impacting $\pol$.
For the choice of offset, we exploit the construction of the model that predicts $q_j$ from atomic positions: To ensure both translational and boundary invariance, the charge at position $j$ depends only on relative positions $(\R_j - \R_i)_{\text{PBC}}$ that point, crucially, \emph{not} neccessarily to the position $i$ located in the unit cell. Rather, $(\R_j - \R_i)_{\text{PBC}}$ always indicates the \emph{closest} replica. In small cells, there may also be multiple relative positions pointing to different replicas of the same original position.
This definition of the BEC is then
\begin{equation}
\label{eq:apt_mic_si}
  \pdv{P_\alpha}{r_{i, \beta}} = q _{\text{IR}}^i \delta_{\alpha, \beta} + \sum_j \pdv{q _{\text{IR}}^j}{r_{i,\beta}} \left(r_{j, \alpha} - r_{i, \alpha}\right)_{\text{PBC}},
\end{equation}
which is boundary invariant. The same expression can be obtained as the $k \rightarrow 0$, i.e., large box, limit of the expression by \citeauthor{zhong25a}~\cite{zhong25a}:
\begin{equation}
    q_i \delta_{\alpha, \beta} + \Re{\sum_j\pdv{q_j}{r_{i, \beta}}  \frac{\exp(ik (r_{j,\alpha} - r_{i, \alpha}))}{ik}} \,.
\end{equation}



\section{Calculating BECs from Learned Charges in Periodic Systems: In Practice}

While \cref{eq:apt_mic_si} is simple in principle, its practical implementation requires additional discussion. This is because as written, it requires explicit access to the partial derivatives $\pdv{q _{\text{IR}}^j}{r_{i,\beta}}$: For each pair of $i$ and $j$, a vector $(\R_j - \R_i)_{\text{PBC}}$ must be assigned. This amounts to evaluating the Jacobian of a multi-valued function (the one that predicts $q_j$) with automatic differentiation, which requires $N$ backward-mode evaluations or $3N$ forward-mode evaluations, neither of which is computationally feasible.

A similar problem occurs in the evaluation of the heat flux for machine-learning interatomic potentials. A solution was presented by \citeauthor{langer23a}~\cite{langer23a,langer23b}, based on two ideas:

First, we do not work in pbc, but in an extended system that explicitly constructs all relevant replica positions.
This construction allows us to split the derivative $\pdv{q_j}{r_{i,\beta}}$ into a sum over all the explicit replicas of $i$ that contribute to $q_j$. Letting a prime index indicate explicitly extended, \enquote{ghost} positions:
\begin{equation}
    \pdv{q_j}{r_{i,\beta}} = \sum_{i'} \pdv{q_j}{r_{i',\beta}} \, .
\end{equation}
Since the $\R_{i'}$ already respect pbc,
\begin{equation}
    (\R_j - \R_{i'})_{\text{PBC}} = \R_j - \R_{i'} \,;
\end{equation}
we no longer require a peculiar measure of the distance between $i$ and $j$.
For local (or semi-local) models, constructing the extended system is feasible as only a small shell of additional positions around the unit cell are required.

Second, we construct an auxiliary quantity $B_\alpha := \sum_j \hat r_{j,\alpha} q_j$ where the positions $\hat r_{j,\alpha}$ are \emph{excluded} from automatic differentiation.
Thus, the gradient of $\vec{B}$ with respect to positions yields
\begin{equation}
    \label{eq:barycenter}
\pdv{B_\alpha}{r_{i',\beta}} = \sum_j r_{j,\alpha} \pdv{q_j}{r_{i',\beta}}.
\end{equation}

Putting it all together, we can write $Z^i_{\alpha,\beta}$ as:
\begin{gather}
    Z^i_{\alpha,\beta} - q_i \delta_{\alpha,\beta} 
      = \sum_j  \pdv{q_j}{r_{i,\beta}} \left(r_{j, \alpha} - r_{i, \alpha}\right)_{\text{PBC}}\\
    = \sum_{j} \sum_{i' } 
      \pdv{q_j}{r_{i',\beta}} \left(r_{j,\alpha} - r_{i',\alpha}\right) \\
    = \sum_{i'} \underbrace{\pdv{B_\alpha}{r_{i',\beta}}} _{\text{3 grad}}
      - \sum_{i'} r_{i',\alpha} 
        \underbrace{\pdv{r_{i',\beta}} \sum_j q_j } _{\text{1 grad}},\label{eq:apt_reform}
\end{gather}
which requires only the evaluation of the gradient of $B_\alpha$ and of $\sum_j q_j$, a total of four backward-mode evaluations with automatic differentiation.

Given that all information needed to explicitly include periodic replicas are inherently given by the neighborlist, implementing \cref{eq:apt_reform} is relatively straightforward.
Thus, we expect an additional factor in the computational cost of the coupled method relative to uncoupled approaches of $\approx 4$.

\begin{figure}
    \includegraphics{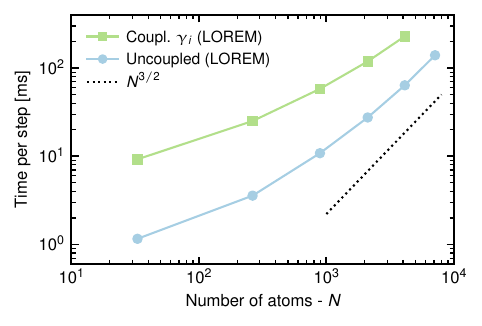}
    \caption{\label{fig:bechmark}
    Benchmark for computational cost of the LOREM models to run a single point evaluation of the energy, force, and BEC for a water structure with varying number of atoms. Each point is an average of 10 single point calculations.}
\end{figure}

This expected factor is visible in the benchmark shown in \cref{fig:bechmark}, where both the coupled and uncoupled architecture approach the expected asymptotical scaling due to the Ewald sum of $\mathcal{O}(N^{3/2})$ with a difference between both architectures that approaches a factor of 3-4 for larger systems.

\section{Analysis of Predictions for the Lorem Architecture}

\begin{figure}
    \includegraphics{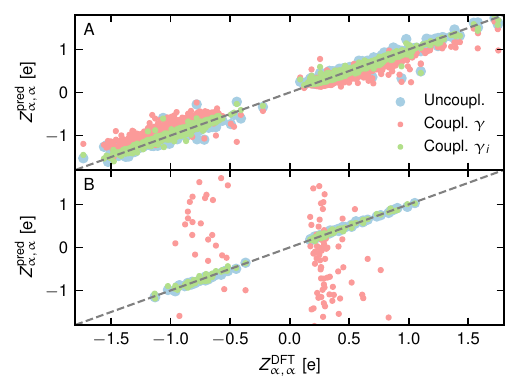}
    \caption{\label{fig:parity_lorem}
        Scatter plot of diagonal components of the BECs comparing DFT labels with model predictions. 
    A shows bulk water and B water clusters.
    }
\end{figure}

Figure \ref{fig:parity_lorem} presents a scatter plot comparing diagonal components of the predicted BEC to DFT labels for bulk and cluster structures, analogous to Fig. 2 in the main text.
The overall trends agree with those reported for the physical architecture: the uncoupled and local $\gamma_i$ model match the DFT labels well, whereas the global $\gamma$ model struggles to simultaneously fit bulk and cluster structures.
For this global screening model, the bulk exhibits slightly smaller errors, whereas the clusters show larger deviations.

\begin{figure}
    \includegraphics{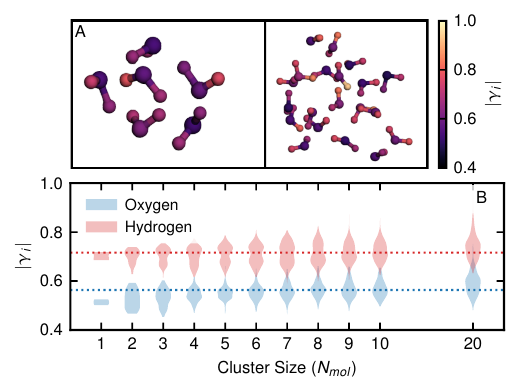}
    \caption{\label{fig:screening_lorem}
    A:~Snapshot of water clusters with 6 and 20 atoms. Atoms are colored according to their local screening values $\gamma_i$. An interactive view is provided online \url{https://doi.org/10.24435/materialscloud:fs-8h}.
    B:~Distribution of the local screening values from the physical coupled $\gamma_i$ model as a function of water cluster sizes. The average bulk values are given by dashed lines.}
\end{figure}

The fitted global $\gamma$ for bulk and cluster structures for a LOREM based global screening coupled model was determined to be $\gamma _{\text{LOREM}} = -0.189$. This value, as well as \cref{fig:screening_lorem} highlight that the LOREM architecture's pseudo charges are no longer physically interpretable, but yield analogous results to the \enquote{physical models} shown in the main text.

\begin{figure}
    \includegraphics{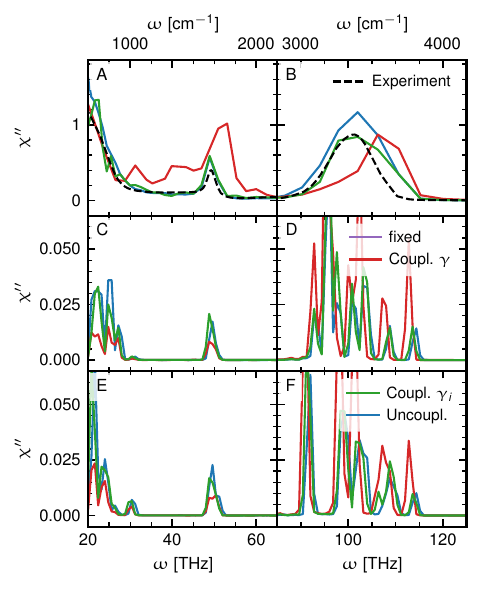}
    \caption{\label{fig:dielectric_spectrum}
      Imaginary part of the complex susceptibility spectrum of periodic bulk water at $T = \SI{300}{\kelvin}$ (A), a cage (B) and book (C) configuration of water hexamer cluster at $T = \SI{10}{\kelvin}$. The gray solid line shows experimental bulk reference spectrum taken from Ref.~\citenum{carlson20a}.
    }
\end{figure}

Figure \ref{fig:dielectric_spectrum} shows---analogous to Fig. 5 in the main text---the imaginary part of the complex susceptibility spectrum for bulk water and two water hexamer cluster configurations, but now calculated with the LOREM architecture. The spectra show the same trends as the physical architectures, with the bulk spectrum matching experimental references well and the cluster spectra showing distinct features depending on the cluster geometry.
Furthermore, we also find the global screening model to agree qualitatively with the other two models (and the experimental spectrum), but to struggle with reproducing the peak heights, which are reproduced more accurately with local screening and uncoupled models.

\section{Stability of Finite Differences Estimates for the BECs}

\begin{figure}
    \includegraphics{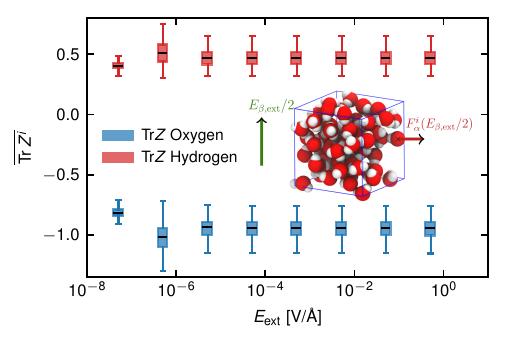}
    \caption{\label{fig:apt_vs_efield}
    Stability of finite differences estimates for the BECs as a function of applied fields $\vec{E}_{\text{ext}}$ in a finite differences scheme. The boxplots show median, first to third quartile (box) and 1.5 times the inter-quartile range (whiskers) of BEC traces in order to give an estimate of the spread in values of the BECs. The schematic inset indicates how the finite field measurements are performed. One can clearly see that estimates are very stable across a wide range of field strengths.}
\end{figure}

Figure \ref{fig:apt_vs_efield} shows the stability of finite differences estimates for the BECs as a function of applied external fields $\vec{E} _{\text{ext}}$ in a finite differences scheme. We find that over almost 9 orders of magnitude of applied field strengths, the estimates for the BECs remain very stable, with only a slight change for very small field strengths due to numerical noise.

